\documentclass[showpacs,superscriptaddress]{revtex4}
\usepackage{graphicx}
\usepackage{amsmath}
\usepackage{amssymb}

\newcommand{\p}{\partial}
\newcommand{\bsigma}{\boldsymbol{\sigma}}
\newcommand{\ph}{\varphi}

\begin{document}

\title{Quasi-bound  states of quantum dots in single and bilayer graphene}
\author{A. Matulis}\email{amatulis@takas.lt}
\affiliation{Departement Fysica, Universiteit Antwerpen \\
Groenenborgerlaan 171, B-2020 Antwerpen, Belgium}
\affiliation{Semiconductor Physics Institute, Go\v{s}tauto 11,
LT-01108 Vilnius, Lithuania}
\author{F.~M.~Peeters}\email{francois.peeters@ua.ac.be}
\affiliation{Departement Fysica, Universiteit Antwerpen \\
Groenenborgerlaan 171, B-2020 Antwerpen, Belgium}

\begin{abstract}
Dirac fermions interacting with a cylindrically symmetric quantum dot potential
created in single and bilayer graphene are not confined but form quasi-bound states.
The broadening of these quasi-bound states (i.~e.~the inverse of their lifetimes)
decreases (increases) with the orbital momentum of the electron
in the case of graphene (bilayer). Quasi-bound states with energy below (above)
the barrier height are dominantly electron(hole)-like. A remarkable decrease
of the energy level broadening is predicted for electron energies close to
the barrier height, which are a consequence of the total internal reflection of
the electronic wave at the dot edge.
\end{abstract}


\date{November 13, 2007}

\maketitle

\twocolumngrid

\section{Introduction}

Quantum dots, or also called ``artificial atoms'', \cite{chak92} are one of the
most intensely studied subjects in present day condensed matter physics.
Recently, the realization of stable single layer and bilayer carbon crystals
(graphene) has aroused considerable interest in the study of their electronic
properties \cite{zh02,nov04,nov05,zh05,gus05}. These new systems exhibit special
excitations which are described by the analogs of the relativistic Dirac
equations \cite{nov06}. One of the most impressive effects is the so called
Klein paradox according to which electrons can cross large electric barriers with
unity probability \cite{kat06,mil07}. As a consequence the control of the electron
behavior by means of electrical potentials and structures becomes a very challenging
task and the creation of quantum dots in such materials is not obvious.

It follows from the Klein effect straightforwardly that the electron will escape
from any potential minimum, and that there are no bound states in an
electrically defined quantum dot (except in gated bilayer\cite{per07}).
However, the polar diagrams presented
in Ref.~\cite{kat06} indicate that in single and bilayer graphene the electron
penetration into a potential barrier is strongly reduced if this electron
propagates at some angles with respect to the barrier. From this observation we
may expect that there might be long living \textit{quasi-bound states} in such
quantum dots for specific orbital momenta of the electrons. These states may be
probed in experiments, say by tunneling currents directed perpendicular to the dot
using e.~g.~STM, or in the near field infrared absorption, as narrow peaks in the
local density of states.

In this work we impose a circularly symmetric quantum dot potential as created
e.~g.~by the split gate technique in single layer and bilayer graphene and discuss
the conditions under which quasi-bounded states can appear.

Such a problem was recently discussed in Ref.~\cite{chen07}
where the semiclassical approach was applied to the problem of electron motion
in a parabolic quantum dot and the imaginary part of the energy
eigenvalues (i.~e.~the lifetime of those states) was calculated. The possibility
to confine the electrons by an external potential in a small region of a graphene
strip was also discussed in Ref.~\cite{silv07}. Our approach is essentially different
from Ref.~\cite{chen07} in two ways: 1) we define the quasi-bound states
through the averaged local density of states, and 2) we study
a quantum dot with a step profile instead of a parabolic dot
where the potential tends to infinity at large distances.
We found that the width of quasi-bound states (i.~e.~the inverse of their lifetime):
1) has the opposite dependence on the angular quantum number for a single and a
bilayer graphene and 2) that it becomes extremely small for energies near the
potential barrier height.

The paper is organized as follows. In section II we formulate the problem of a quantum dot
in graphene. Its solution is given in section III, and in section IV the averaged local
density of states is considered. The results for a single graphene layer are presented in
section V. In section VI the same quantum dot problem is formulated for a bilayer of
graphene, and its solution and results are given in section VII. Our conclusions
are formulated in section VIII.

\section{Quantum dot in graphene}

From the point of view of its electronic properties, graphene can be considered as
a two-dimensional zero-gap semiconductor with its low-energy quasiparticles
(electrons and holes) described by the Dirac-like Hamiltonian \cite{kat06}:
\begin{equation}\label{ham0}
  H_0 = v_F\bsigma{\bf p},
\end{equation}
where $v_F\approx 10^6$ m\,s$^{-1}$ is the Fermi velocity and $\bsigma=(\sigma_x, \sigma_y)$
are the Pauli matrices. We assume that the total Hamiltonian consists of the above
Hamiltonian of free particles in addition to the cylindrically symmetric electric
confinement potential
\begin{equation}\label{pot}
  V({\bf r}) = V\Theta(a-r) = \left\{\begin{array}{ll}
  0, & 0 \leqslant r < a; \\ V, & a \leqslant r < \infty.\end{array}\right.
\end{equation}
Such potential can be created by means of a patterned gate electrode with the
edge smearing much less than the characteristic Fermi wavelength of the electrons and
in turn much larger than the graphene lattice constant.

As the electric potential can not confine the electrons in a finite region of
the graphene plane there are no bound states. Consequently, the
electron will have a continuous spectrum, and electron states in the quantum dot
have to be described as decaying  quasi-bound states. Nevertheless we shall
consider this problem as a stationary one by artificially confining the electron
within a large sample. We shall solve the stationary equation
\begin{equation}\label{sred}
  \{H - E\}\Psi({\bf r}) = 0, \quad H = H_0 + V({\bf r})
\end{equation}
in a finite graphene circle of radius $R$.
In this case the energy spectrum consists of discrete levels separated by
intervals which go to zero as $R^{-1}$ when $R$ tends to infinity.
The presence of quasi-bound states in the quantum dot can show up as peaks in
the averaged electron density in the dot. They can be revealed
experimentally as peaks in the tunneling current through this dot in e.~g.~a
STM experiment.

\section{Solution of eigenvalue problem}

For the sake of convenience we introduce dimensionless variables,
measuring the distance in the units of the quantum dot radius $a$, and
energies in $\hbar v_F/a$ units. For instance, for a dot with radius
$a = 0.1\,\mu$m the above energy unit is of 6 meV. It enables us to
write the Hamiltonian as
\begin{equation}\label{ham}
  H = \begin{pmatrix} V\Theta(1-r) & -i(\p/\p x -i\p/\p y) \\
  -i(\p/\p x +i\p/\p y) & V\Theta(1-r) \end{pmatrix}.
\end{equation}
Lets write the two component wave function as
\begin{equation}\label{wf}
  \Psi({\bf r}) = \begin{pmatrix} A({\bf r}) \\ B({\bf r}) \end{pmatrix},
\end{equation}
and taking into account the cylindrical symmetry of the problem, where we use
\begin{equation}\label{oper}
  \frac{\p}{\p x} - i\frac{\p}{\p y} = e^{-i\ph}\left(\frac{\p}{\p r}
  -\frac{i}{r}\frac{\p}{\p\ph}\right),
\end{equation}
the eigenfunction problem (\ref{sred}) can be written as two coupled differential equations
for the wave function components:
\begin{subequations}\label{twoset}
\begin{eqnarray}
  (V - E)A &=& ie^{-i\ph}\left(\frac{\p}{\p r} - \frac{i}{r}
  \frac{\p}{\p\ph}\right)B, \\
  (V - E)B &=& ie^{i\ph}\left(\frac{\p}{\p r} + \frac{i}{r}
  \frac{\p}{\p\ph}\right)A.
\end{eqnarray}
\end{subequations}
These equations can be further simplified by using the circle symmetry of our
problem which allows us to assume the following angular
dependence of the wave function components:
\begin{equation}\label{angdep}
  \begin{pmatrix} A({\bf r}) \\ B({\bf r}) \end{pmatrix}
  = e^{im\ph}\begin{pmatrix} a(r) \\ ie^{i\ph}b(r) \end{pmatrix},
\end{equation}
where the integer $m$ stands for the eigenstate angular momentum. This
assumption converts Eqs.~(\ref{twoset}) into the following set of coupled ordinary radial
differential equations:
\begin{subequations}\label{twoseto}
\begin{eqnarray}
\label{twoseto1}
  (V - E)a &=& -\left(\frac{d}{dr} + \frac{m+1}{r}\right)b, \\
\label{twoseto2}
  (V - E)b &=& \left(\frac{d}{dr} - \frac{m}{r}\right)a.
\end{eqnarray}
\end{subequations}
These two equations have to be solved in the inner ($r<1$) and outer ($1<r<R$)
region of the dot ensuring the continuity of both wave function components ($a$
and $b$) at the quantum dot edge $r=1$. Moreover, the proper boundary condition
has to be satisfied at the sample edge ($r=R$). Although the exact boundary
condition depends on which sublattice atoms are on the sample edge,
we restrict our consideration to the simple equation
\begin{equation}\label{bca}
  a(R) = 0,
\end{equation}
as the average local density of states which we are looking for is not sensitive
to the microscopic details of the sample edge.

Now inserting $b$ from Eq.~(\ref{twoseto2}) into Eq.~(\ref{twoseto1})
we arrive at the second order ordinary differential equation
\begin{equation}\label{equone}
  \left(\frac{d^2}{dr^2} + \frac{1}{r}\frac{d}{dr} - \frac{m^2}{r^2}\right)a
  = - (V-E)^2a,
\end{equation}
which actually coincides with the Bessel function equation. The other wave
function component can be easily obtained from Eq.~(\ref{twoseto2}).
Thus, the solution inside the dot (where $V(r)=0$) in the case of positive
energy $E>0$ reads
\begin{subequations}\label{solin}
\begin{eqnarray}
  a = FJ_m(Er), \\
  b = FJ_{m+1}(Er).
\end{eqnarray}
\end{subequations}
Note we did not include the Bessel function of second order $Y_m(Er)$ into our
solution, as it is singular at $r=0$.

Outside the dot ($1\leqslant r < R$) the solution is
\begin{subequations}\label{solout}
\begin{eqnarray}
  a &=& PJ_m(\kappa r) + QY_m(\kappa r), \\
  b &=& \mp \left\{PJ_{m+1}(\kappa r) + QY_{m+1}(\kappa r)\right\},
\end{eqnarray}
\end{subequations}
where
\begin{equation}\label{kappa}
  \kappa = |E - V|,
\end{equation}
and the sign in the right hand side of the $b$ expression coincides with the
sign of $(E-V)$.

Satisfying the boundary conditions we obtain the following set of algebraic equations
for the coefficients ($F$, $P$, $Q$):
\begin{subequations}\label{bcs}
\begin{eqnarray}
\label{bcs1}
&&  FJ_m(E) = PJ_m(\kappa) + QY_m(\kappa), \\
\label{bcs2}
&&  FJ_{m+1}(E) = \mp\left\{PJ_{m+1}(\kappa) + QY_{m+1}(\kappa)\right\}, \\
\label{bcs3}
&&  PJ_m(\kappa R) + QY_m(\kappa R) = 0.
\end{eqnarray}
\end{subequations}

As we are interested in the limiting case $R\to\infty$, we replace the Bessel
functions in Eq.~(\ref{bcs3}) by their asymptotic expressions, which results into
\begin{equation}\label{third}
  P\cos\left(\kappa R - \frac{\pi m}{2} - \frac{\pi}{4}\right)
  + Q\sin\left(\kappa R - \frac{\pi m}{2} - \frac{\pi}{4}\right) = 0.
\end{equation}
Up to a normalization factor the solution of this equation can be
chosen as
\begin{subequations}\label{solnorm}
\begin{eqnarray}
  P &=& \sin\left(\kappa R - \frac{\pi m}{2} - \frac{\pi}{4}\right), \\
  Q &=& -\cos\left(\kappa R - \frac{\pi m}{2} - \frac{\pi}{4}\right).
\end{eqnarray}
\end{subequations}

\section{Local density of states}
\label{sec_lds}

Inserting the obtained solution for the $P$ and $Q$ coefficients into
Eqs.~(\ref{bcs}a,b) we arrive at the equation for the eigenvalues.
It follows from Eqs.~(\ref{solnorm}) that these eigenvalues are separated by
\begin{equation}\label{interval}
  \Delta E = \Delta\kappa = \frac{\pi}{R}.
\end{equation}
These dense discrete energy levels are a consequence of the finite size of our
sample and are not of interest to us. Therefore we choose
another procedure and solve the two equations (\ref{bcs}a,b)
together with the condition
\begin{equation}\label{conpq}
  P^2 + Q^2 = 1
\end{equation}
which is clearly satisfied as it follows from Eq.~(\ref{solnorm}).
This procedure enables us to obtain the local density of states from which we can
derive the quasi-bound states. The procedure is as follows.

First, we calculate
the normalization factor $N$ of the wave function.
In the limiting case $R\to\infty$ it can be done just using the asymptotic
wave function expression, namely,
\begin{equation}\label{normfact}
\begin{split}
& 1 = 2\cdot 2\pi N^2\int_0^Rrdr\left(\frac{2}{\pi\kappa r}\right)\Big\{
  P\cos\left(\kappa r - \frac{\pi m}{2} - \frac{\pi}{4}\right) \\
&  + Q\sin\left(\kappa r - \frac{\pi m}{2} - \frac{\pi}{4}\right)\Big\}^2
  = \frac{4N^2R}{\kappa}\left(P^2 + Q^2\right).
\end{split}
\end{equation}
The additional factor $2$ appears because both wave function components have
to be taken into account. Using Eq.~(\ref{conpq}) we find
\begin{equation}\label{nf}
  N = \sqrt{\frac{\kappa}{4R}}.
\end{equation}

Next, we point out that any physical property of the quantum dot,
say like the tunneling current through the dot, or the  absorption of the infrared
radiation in near field spectroscopy, can be expressed as a summation of some matrix
elements over the above dense quantum states. The matrix elements are integrals over
the quantum dot area, namely,
\begin{equation}\label{matelem}
  M = 2\pi N^2F^2\int_0^1rdrf(r)\left\{J_m^2(Er) + J_{m+1}^2(Er)\right\},
\end{equation}
where the function $f(r)$ characterizes the interaction of the quantum dot
with the measuring probe. Thus, replacing the summation over the discrete levels
by an integration, in accordance with Eq.~(\ref{interval}),
\begin{equation}\label{sumint}
  \sum_nM \approx \frac{R}{\pi}\int dEM \sim 2R\int dE N^2F^2
  = \int dE \frac{\kappa F^2}{2},
\end{equation}
we see that within the accuracy of the experimental formfactor (the integral
in the right hand side of Eq.(\ref{matelem})) the quantity
\begin{equation}\label{dens}
  \rho(E) = \frac{1}{2}|E - V|F^2
\end{equation}
acts as a local density of states in the quantum dot area.

\section{Results for quantum dot in graphene}

\begin{figure}[h]
\begin{center}\leavevmode
\includegraphics[width=8cm]{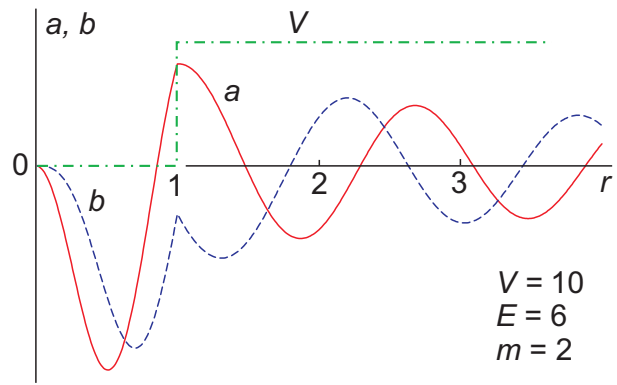}
\caption{(Color on line) Profile of the confinement potential (dot-dash curve),
and the two wave function components: $a$ -- solid curve,
$b$ - dashed curve. Barrier height $V = 10$, energy $E = 6$,
and orbital momentum $m = 2$.}
\label{fig1}
\end{center}
\end{figure}

We solved Eqs.~(\ref{bcs}a,b,\ref{conpq}) numerically from which we obtained the
coefficients, and the local density of states (\ref{dens}).
A typical example for the two components of the wave function together with the
confinement potential profile is shown in Fig.~\ref{fig1}.

As the energy is lower than the potential height we see that the two
wave function components have a different phase indicating the electronic type
character of the wave function inside the dot, and the hole type character outside it.
The large value of the wave function components inside the dot show that this
eigenfunction is a quasi-bound state.

\begin{figure}[h]
\begin{center}\leavevmode
\includegraphics[width=8cm]{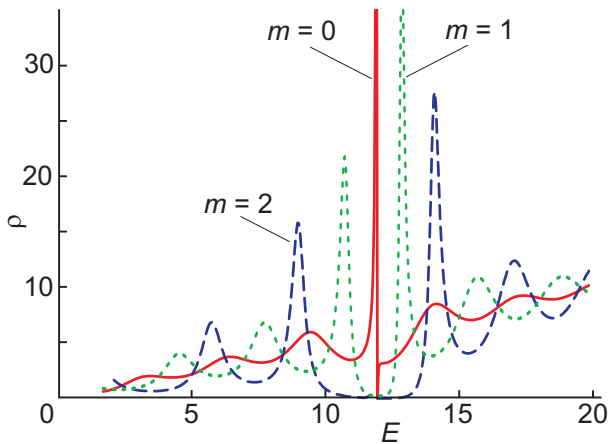}
\caption{(Color on line) Local density of states as function of the energy in case of
barrier height $V = 12$ for the orbital momenta $m = 0$ (solid curve), $m = 1$ (dotted curve),
and $m = 2$ (dashed curve).}
\label{fig2}
\end{center}
\end{figure}

\begin{figure}[h]
\begin{center}\leavevmode
\includegraphics[width=8cm]{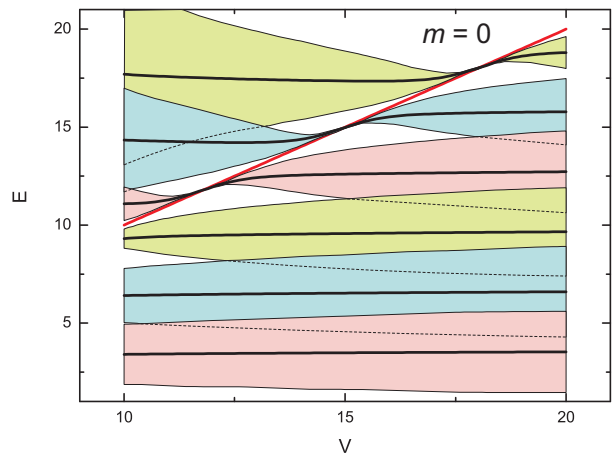}
\caption{(Color on line) Quasi-bound states with orbital momentum $m = 0$
for a quantum dot in graphene. The energy of these states are given by
the black curves and its width (i.~e.~the inverse of the life time) by the
colored region. The straight slanted line corresponds to $E=V$.}
\label{fig3}
\end{center}
\end{figure}

\begin{figure}[h]
\begin{center}\leavevmode
\includegraphics[width=8cm]{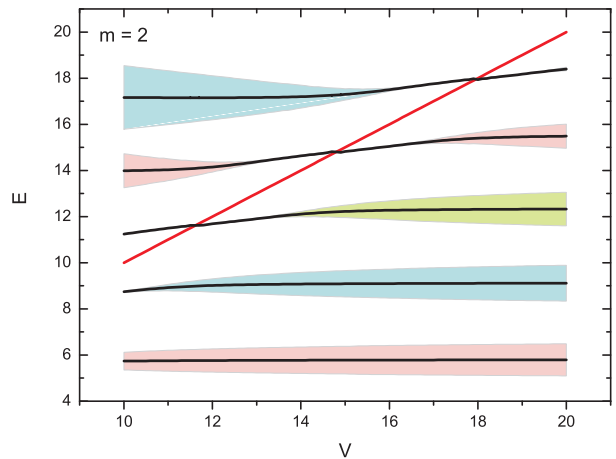}
\caption{(Color on line) The same as Fig.~\ref{fig3} but now for $m = 2$.}
\label{fig4}
\end{center}
\end{figure}

Typical local density of states are shown in Fig.~\ref{fig2}.
It exhibits peaks which can be associated with the quasi-bound states
of the dot. The three curves correspond to the following orbital momenta of electron
$m = 0,1,2$. We observe the general tendency that the larger the orbital
momentum the narrower the peaks. Noticeable very narrow peak
when the energy is close to the barrier height (see the curve for $m = 0$).
This tendency is even better seen in Figs.~\ref{fig3} and \ref{fig4} where the
positions and broadenings of the peaks are shown.

We fitted peaks in the density of states by Lorentzian functions
$a_n\gamma_n/\{(E-E_n)^2+\gamma_n^2\}$ defining three parameters for any of them:
the position $E_n$, its broadening $\gamma_n$, and the amplitude $a_n$.
Graphically these parameters are shown in Figs.~\ref{fig3} and \ref{fig4}
for two orbital momentum $m$ values as function of the barrier height $V$.
The positions $E_n$ of the quasi-bound levels are shown by the solid curves
while the shaded areas between two $E_n\pm\gamma_n$ curves indicate the broadening
of them.

As expected, in the case $m = 0$ the levels are rather broad. Actually, they
can hardly be identified as quasi-bound levels and they rather correspond to weak
oscillations in the local density of states of the continuous spectrum
(see the solid curve in Fig.~\ref{fig2}).

In the case of $m=2$ we see (Fig.~\ref{fig4}) a quite  different picture. The
levels are narrow and indicate the presence of long living quasi-bound states.
It is interesting to see that the quasi-bound states are seen above as well
as below the barrier, the latter is indicated by the slant solid line. Actually, this is
the consequence of the equivalence of the Dirac electrons and holes in the barrier region.

\begin{figure}[h]
\begin{center}\leavevmode
\includegraphics[width=6cm]{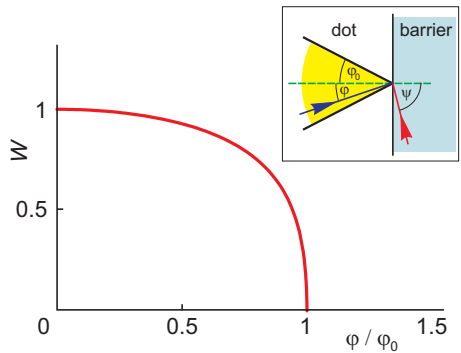}
\caption{(Color on line) Tunneling probability dependence on the incident angle of
the electron beam in the case of small deviation of the electron energy from
the barrier height. Inset: definition of the critical angle $\ph_0$.}
\label{fig5}
\end{center}
\end{figure}

As was already mentioned in Figs.~\ref{fig3} and \ref{fig4} we see one
more important peculiarity of the local density of states of a quantum dot in graphene.
This is the extremely narrow states in the vicinity of the top of the barrier.
This is not an accidental phenomena, but the consequence of the important fact
that it is rather difficult for Dirac electrons to penetrate the barrier when its
energy is close to the barrier height.
This property follows straightforwardly from the electron penetration
through the barrier problem solved in \cite{kat06,mil07}, although they payed no attention
to this limit case. The matter is that the angle $\ph$ (with respect to the
perpendicular to the barrier, see the inset in Fig.~\ref{fig5}) of the incident
electron and the angle $\psi$ of the refracted electron has to satisfy the equation
\begin{equation}\label{sinchr}
  E\sin\ph = (V - E)\sin\psi,
\end{equation}
which is the equivalent of Snell's law in optics \cite{che07}.
In the case when the electron energy is close to the barrier height
\begin{equation}\label{bv}
  E = V - \Delta, \quad |\Delta| \ll V,
\end{equation}
the electron wave goes from the material with large refraction index into
the material with small refraction index. In this case the well known phenomena of
total internal reflection takes place. It means that there is a critical incident angle
\begin{equation}\label{critang}
  \ph_0 = |\Delta| / V,
\end{equation}
such that electrons with larger incident angles ($|\ph| > \ph_0$) are totally reflected
from the barrier (see the inset in Fig.~\ref{fig5}, where the angles at which the
electron penetrates the barrier are shown by the shadowed sector). The electron
current (which can be named as tunneling probability) in the barrier perpendicular
to the barrier edge direction can be expressed as follows:
\begin{equation}\label{tunsr}
  W = \frac{2\sqrt{\ph_0^2 - \ph^2}\,\Theta(\ph_0 - \ph)}
  {\ph_0 + \sqrt{\ph_0^2 - \ph^2}},
\end{equation}
and is shown in Fig.~\ref{fig5}.

So, when the electron energy is close to the barrier height there is
only a very small region of the incident angles at which the electron can penetrate
the barrier. It means that for these electrons the electrical barrier acts
almost as a confining potential. This explains why in this energy region
very narrow quasi-bound levels appear even for zero orbital momentum.

\section{Quantum dot in bilayer}

It is instructive to consider the same quantum dot problem for bilayer graphene.
There are similarities and differences between the two graphene systems \cite{par06}.
Both systems have gapless electron and hole spectra, but are described
by different Dirac type Hamiltonians.

Following Ref.~\cite{kat06} we use the two component wave function approximation
for the bilayer which is described by the following Hamiltonian:
\begin{equation}\label{bham}
  H_0 = \frac{1}{2}\begin{pmatrix} 0 & (\p/\p x - i\p/\p y)^2 \\
  (\p/\p x + i\p/\p y)^2 & 0\end{pmatrix},
\end{equation}
where the distances are measured as before in units of the quantum dot radius $a$,
and the energies are measured in $\hbar^2/m^*a^2$ units. For example, for a dot with
radius $a = 0.1\,\mu$m, and the effective mass \cite{par06} $m^*=0.038 m_e$ the
above energy unit is $0.2$ meV.

Assuming expressions for the wave function analogous to those for graphene (\ref{wf})
we obtain the following set of equations for the wave function components:
\begin{subequations}\label{eqcomp0}
\begin{eqnarray}
\label{eqcomp01}
  (E - V)A &=& \frac{1}{2}\left(\frac{\p}{\p x} - i\frac{\p}{\p y}\right)^2B, \\
\label{eqcomp02}
  (E - V)B &=& \frac{1}{2}\left(\frac{\p}{\p x} + i\frac{\p}{\p y}\right)^2A.
\end{eqnarray}
\end{subequations}
The components $A$ and $B$ have to satisfy the continuity condition together with
continuity of their first radial derivatives at the dot edge.

In the dot and as well outside it these two equations can be transformed
into a single equation for any wave function component:
\begin{equation}\label{compfact}
\begin{split}
&  \left\{\frac{1}{2}\nabla^4 - (E-V)^2\right\}A \\
&  = \left\{\frac{1}{2}\nabla^2 + (E-V)\right\}
  \left\{\frac{1}{2}\nabla^2 - (E-V)\right\}A = 0.
\end{split}
\end{equation}
Using the axial symmetry and assuming the following angular dependence:
\begin{equation}\label{bbf}
  A({\bf r}) = e^{im\ph}a(r),
\end{equation}
this component has to satisfy any of the following radial equations:
\begin{equation}\label{blgcomp}
  \left\{\frac{d^2}{d r^2} + \frac{1}{r}\frac{d}{d r}
  + \left[\pm (E-V) - \frac{m^2}{r^2}\right]\right\}a = 0.
\end{equation}
It is evident that they are the equations for the Bessel and modified Bessel functions.
Inside the quantum dot ($r \leqslant 1$) the solution is:
\begin{equation}\label{btv}
  a = FJ_m(kr) + GI_m(kr),
\end{equation}
where $k = \sqrt{2E}$. Two other functions ($Y_m$ and $K_m$) are not included
because of their singularity at the origin $r=0$. Similarly, using Eq.~(\ref{eqcomp02})
we obtain for the other component of the wave function
\begin{subequations}\label{bcompb}
\begin{eqnarray}
  B({\bf r}) &=& e^{i(m+2)\ph}b(r), \\
  b &=& FJ_{m+2}(kr) + GI_{m+2}(kr).
\end{eqnarray}
\end{subequations}

Outside the quantum dot the wave function components are given by
\begin{subequations}\label{boutdot}
\begin{eqnarray}
  A &=& PJ_m(\kappa r) + QY_m(\kappa r) + SK_m(\kappa r), \\
  B &=& \mp\big\{PJ_{m+2}(\kappa r) + QY_{m+2}(\kappa r) \nonumber \\
  &&\phantom{mm} + SK_{m+2}(\kappa r)\big\},
\end{eqnarray}
\end{subequations}
where $\kappa = \sqrt{2|E - V|}$, and the sign of the $B$ component coincides
with the sign of the expression $(E - V)$. We do not included the functions
$I_n$ and $I_{m+2}$ into the above expressions, as these functions are
responsible only for satisfying the boundary condition at the sample edge
$r = R$, and are therefore not relevant in the limit $R\to\infty$.

In analogy to our approach for graphene we equate both
wave function components and their derivatives at the quantum dot edge ($r = 1$),
and together with Eq.~(\ref{conpq}) we obtain a set of five algebraic equations
for the five parameters $F$, $G$, $P$, $Q$ and $S$.

We point out that the functions $I_m(kr)$ and $K_m(\kappa r)$ together with the
analogs for orbital momentum $m+2$ are essential only in the region close to the
dot edge, where they ensure the continuity of the derivatives of the wave
function components. While the main contribution to the local density of states
is determined by the functions $J_m(kr)$ and $J_{m+2}(kr)$. Following the
procedure presented in Sec.~\ref{sec_lds} we obtain the following expression
for the quantum dot local density of states in a bilayer:
\begin{equation}\label{bdens}
  \rho(E) = \frac{1}{2}F^2.
\end{equation}
Note that it differs from the analogous expression for the case of graphene by the
factor depending on the electron energy in the barrier, what is caused by
the different dispersion law in both materials.

\section{Results for the dot in bilayer}

\begin{figure}[h]
\begin{center}\leavevmode
\includegraphics[width=8cm]{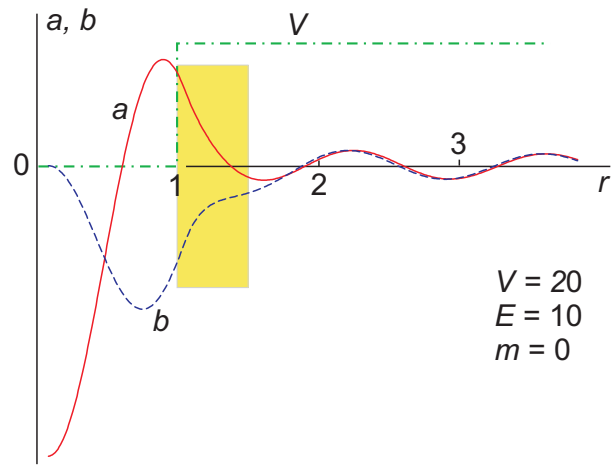}
\caption{(Color on line) Wave function components in bilayer: $V = 20$, $E = 10$,
$m=0$: solid curve -- component $a$, dashed curve -- component $b$,
the dash-dotted curve is the potential profile of the dot.}
\label{fig6}
\end{center}
\end{figure}

Solving numerically the set equations for the coefficients, introduced in the previous
section, we calculated the wave functions and the local density of states.
An example of the wave function is shown in Fig.~\ref{fig6}.
Comparing it with the wave function for the dot in graphene we notice several
differences. The derivative of the wave function in a bilayer is continuous at
the dot edge, while in graphene the wave function components exhibit
kinks there. That is the reason why in a bilayer there is an intermediate region of
exponential behavior (see the shadowed rectangle in Fig.~\ref{fig6})
where the electron type function changes itself into the hole type one.

\begin{figure}[h]
\begin{center}\leavevmode
\includegraphics[width=8cm]{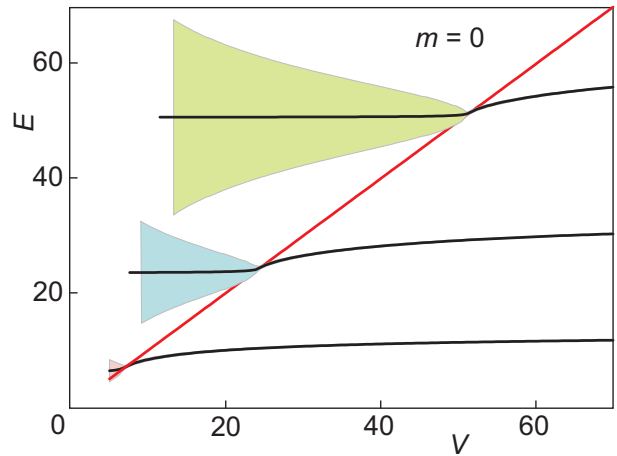}
\caption{(Color on line) Quasi-bound states for quantum dot in bilayer with
orbital momentum $m = 0$. The energy of these states are given by
the black curves and its width (i.~e.~the inverse of the life time) by the
shadowed regions. The straight slanted line corresponds to $E=V$.}
\label{fig7}
\end{center}
\end{figure}

\begin{figure}[h]
\begin{center}\leavevmode
\includegraphics[width=8cm]{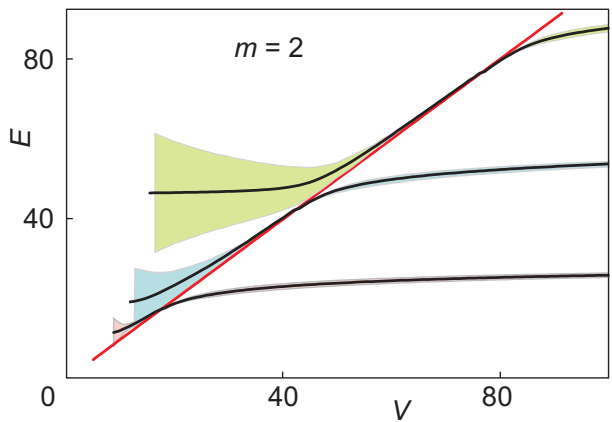}
\caption{(Color on line) The same as Fig.~\ref{fig7} but now for $m = 2$.}
\label{fig8}
\end{center}
\end{figure}

The local density of states of the quantum dot states is shown in Figs.~\ref{fig7} and
\ref{fig8} in the case of two orbital momenta ($m=0$ and m=2).
These pictures also differ essentially from those for graphene shown in
Figs.~\ref{fig3} and \ref{fig4}.
First, we see that the quasi-bound states in a bilayer are much
narrower as compared with these in graphene. The main reason of this difference
is as follows. Although single layer graphene and bilayer graphene are both gapless
materials, the physical nature of their bands is quite different.
In graphene the electron and hole parts of the bands are the natural
prolongation of each other. Electrons and holes are like different
expressions of the same Dirac quasi-particle. While in bilayer graphene these two
contiguous electron and hole bands are much more independent from one
another, and reminds one to the accidental touch of their extremum points.

This can be more clearly demonstrated by rewriting the component equations
(\ref{eqcomp0}) in the case of perpendicular incidence of the electron to the barrier:
\begin{subequations}\label{eqcompp}
\begin{eqnarray}
\label{eqcompb1}
  (E - V)A &=& \frac{1}{2}\frac{d^2}{dx^2}B, \\
\label{eqcompb2}
  (E - V)B &=& \frac{1}{2}\frac{d^2}{dx^2}A.
\end{eqnarray}
\end{subequations}
Adding and subtracting these equations we obtain the uncoupled equation set
for the hole $A+B$ and electron $A-B$ type wave functions.
Consequently, in the case of perpendicular incidence the electron in a bilayer
can be confined in the dot by the electrical potential, and thus, we obtain
stationary states. In the case of slanted incidence, when the electron momentum
component $k_{\parallel}$ along the barrier is not zero the equations
can no longer be decoupled. In this case this longitudinal momentum (or
the orbital momentum of electron) serves as a coupling constant between the electron
and the hole. Consequently, the larger the angular momentum, the larger the
probability for the electron to convert itself into a hole, or the more dominant
is the Klein effect, and thus the more smeared is the
quasi-bound level. This is actually seen in Figs.~\ref{fig7} and \ref{fig8}
which show that the quasi-bound states for $m = 0$ are narrower as
compared to the $m=2$ states.
By the way, this difficult penetration of the electron into the barrier is indicated
by the intermediate exponential region in the wave function
coordinate dependence as shown by the shadowed area in Fig.~\ref{fig6}.

This simple physical picture explains one more interesting property of the
above local dot density of states.
In Figs.~\ref{fig3} and \ref{fig4} we see that in graphene there is some symmetry
between the quasi-bound states below the top of the barrier (blue solid
slanted line) and above it.
While in bilayer graphene as seen in Figs.~\ref{fig7} and \ref{fig8}
such symmetry is absent and the states above and below the top are quite different.
The states above the top of the barrier are much more smeared. This is
caused by the fact that above the barrier the nature of the wave function is
the same in both regions (in the dot and outside it), and there is no need
for the electron to transform itself into a hole, and consequently, the
probability to escape the dot is larger than in the opposite case when
the energy is smaller than the top of the barrier.

\section{Summary and conclusions}

Using the two wave function component approximation we calculated the local density
of states in an electrically defined circle symmetric quantum dot in single layer and
bilayer graphene. It was shown that in bilayer rather narrow quasi-bound states
appear when the energy is smaller than the barrier height. The broadening of the
states in bilayer graphene increases as the orbital momentum becomes larger
which is opposite to the case of graphene.

In contrast in graphene narrow quasi-bound states are predicted with
increasing orbital momentum. This different behavior of the quasi-bound
states in graphene and bilayer is explained by the different physical nature of
the touching electron and hole energy bands.

Weakly broadened quasi-bound states are predicted in both graphene and bilayer graphene
in the region where the electron energy is close to the top of the barrier.
This phenomena can be understood from an analog of the optical effect of total internal
reflection which an electron wave suffers in the above mentioned region of energies.

We also notice the different symmetry of the states above and
below the top of the barrier in single layer graphene and bilayer graphene, caused by
the different way of electron conversion into a hole in both systems.

\begin{acknowledgments}
This work is supported by the Flemish Science Foundation (FW0-Vl),
the Europian Network of excellence SANDiE, and the Interuniversity
attraction poles programme (IAP), Belgian Science Policy, Belgian State.
\end{acknowledgments}

\onecolumngrid

\end{document}